\documentclass[twocolumn,pre,preprintnumbers,amsmath,amssymb]{revtex4}
\usepackage{amssymb,epsf}
\usepackage{latexsym}                                  
\usepackage{graphicx}
\usepackage{float}
\begin{document}

\title{Dynamic Black Holes in a FRW background: Lemaitre transformations}
\author{H. Moradpour\footnote{h.moradpour@riaam.ac.ir}, A. Dehghani,
M. T. Mohammadi Sabet\footnote{mohammadi.sabet@gmail.com}}
\address{Research Institute for Astronomy and Astrophysics of Maragha (RIAAM),
P.O. Box 55134-441, Maragha, Iran.\\
}

\begin{abstract}
Since the conformal transformations of metric do not change its causal structure,
we use these transformations to embed the Lemaitre metrics into the FRW background. In
our approach, conformal transformation is in agreement with the universe expansion
regimes. Indeed, we use the Lemaitre metrics because the horizon singularity is
eliminated in these metrics. For our solutions, there is an event horizon while its
physical radii is increasing as a function of the universe expansion provided suitable
metrics for investigating the effects of the universe expansion on the Black Holes.
In addition, the physical and mathematical properties of the introduced metrics have
well-defined behavior on the event horizon. Moreover, some physical and mathematical
properties of the introduced metrics have been addressed.

\end{abstract}

\maketitle

\section{Introduction}
The universe expansion is described by Friedmann-Robertson-Walker
(FRW) spacetime,
\begin{eqnarray}\label{FRW}
ds^2=-dt^2+a(t)^2[\frac{dr^2}{1-kr^2}+r^2d\Omega^2],
\end{eqnarray}
where, $d\Omega$ is the line element on the unit $2$-sphere
hypersurface and $k$ is the background curvature which points to flat,
open and close universe for $0$, $-1$ and $1$, respectively. According
to the standard cosmology, universe can be expanded either as a power
law $a(t)=At^{\frac{2}{3(\omega+1)}}$ for $\omega>-1$ or exponentially
$a(t)=A\exp{Ht}$ for $\omega=-1$ (dark energy), depending on the equation
of the state parameter $\omega=\frac{P}{\rho}$. Furthermore, for $\omega<-1$,
known as Phantom dark energy, the scale factor increases as
$a(t)=A(t_{br}-t)^{\frac{2}{3(\omega+1)}}$. In the above statement,
$H(\equiv\frac{\dot{a}}{a})$ and $t_{br}$ are the Hubble parameter and the Big
Rip singularity time, when the expansion of universe ends
catastrophically and everything will ultimately decompose into its
elementary constituents, respectively \cite{mukh}. Some aspects of the conformal
form of the FRW metric has been studied in Refs. \cite{rma,iloshi}.
In this metric $\xi=a(t)r$ is the physical radii in this metric and
$r$ is called co-moving radius. It is be mentioned that observations support the flat
universe ($k=0$) \cite{wein}.

In order to find the effects of the universe expansion on local
physics, McVittie derived a metric explaining a point-like
chargeless particle in the FRW background \cite{mcvittie}. considering
the perfect fluid concept together with the isotropic form of the
FRW metric, the Mcvittie spacetime can be generalized to
arbitrary dimensions and charged objects \cite{gao,gao1}. It is
easy to show that the curvature scalars diverge at the redshift
singularity and energy conditions are not satisfied \cite{md}.
Also, the radii of the redshift singularity depends on the background
curvature ($k$) and the mass and the charge are decreased by the
expansion \cite{gao1}. These failures are the unsatisfactory parts
of these attempts and therefore, they may not provide a suitable metric for
studying dynamical Black Holes (BHs) \cite{md,nolan,nolan1,nolan2,susman,feraris}.

Using perfect fluid concept in a non-static spherically symmetric
background and and considering the universe expansion, one can
find solutions including constant mass, charge and cosmological constant \cite{mr}.
On one hand, the curvature scalars diverge on the redshift singularity
as the Mcvittie spacetime. On the other hand, the redshift singularity
is independent of the background curvature contrary to the Mcvittie
solution and its generalizations. In addition, The first law
of thermodynamics is satisfied on hypersurface with radii equal to the
redshift singularity. In continue and just same as the Mcvittie spacetime,
energy conditions are not satisfied near this hypersurface.
Finally, like the Mcvittie spacetime and its
generalizations, it seems that these solutions are not suitable
for studying dynamical BHs. Indeed, these two class of solutions
may contain naked singularity which attracted more attempts to itself,
since it was shown that the naked singularity can be considered
as a possible source for gravitational lansing in astrophysical
observations \cite{mr,virb,virb1}. More studies in which the prefect fluid
concept is used to derive the dynamic spherically symmetric solutions
can be found in Refs. \cite{rosen1,rosen2}.

In another approach, Einstein et al. tried to connect the
Schwarzschild spacetime to the FRW spacetime on a timelike
hypersurface as a boundary \cite{es}. In this model the
Schwarzschild spacetime explains the inner spacetime where the FRW
metric is used to describe the outer spacetime while junction
conditions should be satisfied on the boundary
\cite{saida,peebl,lake,fleu,kan1,roed,kan}. Since the inner
spacetime is illustrated by the Schwarzschild metric, these
solutions do not include dynamical BHs. Moreover, these solutions
satisfy energy conditions and are classified into a more general
class of solutions named Swiss-Cheese models \cite{md}.

It is shown that the conformal transformation does not change the
causal structure of the spacetime \cite{wald}. Therefore, it arouses
considerable interests in order to find dynamical BHs. Inspired by
this property of the conformal transformation, some authors have
considered the Schwarzschild and Reissner-Nordstrom metrics in
various coordinates and then they tried to lay these BHs into the
FRW background by using conformal transformation while conformal
factor explains the universe expansion \cite{md,vai,tha,dy1}.
Indeed, this technique helps us find new metrics. Thakurta spacetime
and its generalization to charged BHs, which include conformal
Schwarzschild and Reissner-Nordstrom metrics in their original
coordinates, can be classified into the one class of solutions that
has gravitational lagrangian (Ricci scalar) conformal to FRW's
\cite{rms}. Indeed, these solutions can be again classified into the
more general class of solutions with special symmetry and including
conformal de-Sitter solutions \cite{mr,mr1}. For these solutions,
curvature scalars diverge at the redshift singularities which are
null hypersurface and they are expanding with time. The
generalization of the Thakurta spacetime to charged BHs does not
satisfy energy condition where the Thakurta spacetime satisfies the
energy conditions out of the redshift singularity \cite{rms}. Since,
the Thakurta spacetime includes the conformal transformation of the
Schwarzschild spacetime and satisfies energy conditions out of the
redshift singularity, it is now accepted that this redshift
singularity may point to a dynamical BH \cite{md,rms}. It was also
shown that the first law of thermodynamics is satisfied on the
horizon of the Thakurta spacetime \cite{fara}. Apparent horizon and
its properties including temperature and the amount of energy which
is confined to it can be found in Ref. \cite{rms}.

There is also a solution introduced by Sultana et al., which includes
conformal transformation of the Eddington-Finkelstein form of the
Schwarzschild metric. This solution does not satisfy energy conditions
everywhere out of the redshift singularity \cite{md}, but
curvature scalars do not diverge at the redshift singularity
which make good sense about the nature of this singularity.
Because of the behavior of the curvature scalars and since this
spacetime is just a conformal transformation of the
Eddington-Finkelstein form of the Schwarzschild metric, it is now
accepted that this spacetime may include dynamical BH
\cite{md,dy1}.

Another solution has been introduced by M$^{\textmd{c}}$Clure et al.,
 which includes conformal transformation of the isotropic form of Schwarzschild
satisfying the energy conditions everywhere \cite{md}. The
generalization of this solution to the charged spacetime has also
been investigated \cite{md}. It is easy to check that the curvature
scalars diverge at the redshift singularity, where is a null
expanding hypersurface, which it seems unsatisfactory. Since these
objects are conformal transformations of isotropic form of the
Schwarzschild and Riessner-Nordstorm metrics, they may point to
dynamical BHs \cite{md}.

In this paper, we firstly review some properties of Lemaitre
metrics. Then, in order to study the effects of the universe
expansion on a BH, we introduce conformal incoming Lemaitre
spacetime and verify its physical and mathematical properties
including redshift, the behavior of the curvature scalars and energy
conditions, in section $\textmd{II}$. In section $\textmd{III}$, we
use the conformal incoming Lemaitre spacetime to investigate the
effects of the universe expansion on the horizon. In addition, we
investigate its physical and mathematical properties. Section
$\textmd{IV}$ is devoted to summery and conclusion.

Throughout this paper, the metric signature we shall adopt is
$(+---)$, and we take $\hbar=c=G=1$ for simplicity.
\section{Conformally inward Lemaitre Spacetime}
The Schwarzschild line element as the stationary vacuum
solution of the Einstein field equations with spherically
symmetry can be written as
\begin{equation}\label{Schwarzchild metric}
dS_S^2 = \left( {1 - \frac{{{r_S}}}{r}} \right)d{t^2} - {\left( {1
- \frac{{{r_S}}}{r}} \right)^{ - 1}}d{r^2} - {r^2}d{\Omega ^2},
\end{equation}
where $d{\Omega ^2} = d{\theta ^2} + \sin\theta^2 d{\phi ^2}$ and
${r_S} = 2M$ are the line element on the unit $2$-sphere and the
Schwarzschild radius, respectively \cite{P}. In addition, $M$ is the
mass enclosed by radii $r_S$, and Schwarzschild metric is only valid
for $r > {r_S}$ while it is also undefined at horizon ($r = {r_S}$).
In order to eliminate the metric singularity at the Schwarzschild
radius one can apply the following coordinate transformations on the
Schwarzschild coordinates $(r,t)$  \cite{lemaitre}
\begin{equation}\label{Lemaitre transformations}
\begin{array}{l}
\bar t = \pm t \pm \int {\frac{{\sqrt {{r_s}/r} }}{{\left( {1 - {r_s}/r} \right)}}} dr,\\
\bar r = t + \int {\frac{{\sqrt {r/{r_s}} }}{{\left( {1 - {r_s}/r}
\right)}}} dr.
\end{array}
\end{equation}
These transformations are called the Lemaitre transformations
eliminating the metric singularity at the Schwarzschild radii and
therefore, the Lemaitre metrics are regular at the event horizon. It
means that these metrics can be used to explain a particle moving
freely in a given field \cite{Shank}. Choosing the positive sign in
\eqref{Lemaitre transformations}, the Lemaitre element for the
freely falling particles becomes
\begin{equation}\label{Lemaitre metric}
dS_L^2 = d{\bar t^2} - \frac{{{r_S}}}{r}d{\bar r^2} -
{r^2}d{\Omega ^2},
\end{equation}
where
\begin{equation}\label{r}
r = {\left( {\frac{3}{2}(\bar r - \bar t)}
\right)^{\frac{2}{3}}}r_S^{\frac{1}{3}}.
\end{equation}
We should note that the metric (\ref{Lemaitre metric}) is
non-stationary while the coordinates $\bar r$ and $\bar t$ are
everywhere spacelike and timelike respectively. It is apparent from
Eq.\eqref{r} that $\bar r - \bar t>0$ must be valid everywhere
\cite{Shank}. $\bar t$ is the proper time for particles which are at
rest in the Lemaitre coordinate system. In these coordinates, the
singularity on the Schwarzschild coordinates corresponds on the
equality ${r_S} = \frac{3}{2}(\bar r - \bar t)$, and particles
cannot remain at rest for $r<r_S$ \cite{Shank,landau}. In addition,
calculations for the Kretschmann scalar lead to ${R^{\mu \nu \rho
\sigma}}{R_{\mu \nu \rho \sigma }} = \frac{64}{27(\bar r - \bar
t)^4}$ showing that there remains an essential curvature singularity
at the origin (where $\bar r - \bar t=0$) like the Schwarzschild
coordinates \cite{landau,lemaitre}. After some algebra we have,
\begin{equation}\label{Lemaitre red shift}
1 + z = \frac{{{\lambda _O}}}{{{\lambda _E}}} = \sqrt {\frac{{1 -
{{\left( {\frac{{2{r_S}}}{{3{{(\bar r - \bar t)}_O}}}}
\right)}^{2/3}}}}{{1 - {{\left( {\frac{{2{r_S}}}{{3{{(\bar r -
\bar t)}_E}}}} \right)}^{2/3}}}}},
\end{equation}
as the redshift of incoming photon at radii $r_O=(\frac{3}{2}(\bar r
- \bar t)_O)^{\frac{2}{3}}r_S^{\frac{1}{3}}$ emitted at
$r_E=(\frac{3}{2}(\bar r - \bar
t)_E)^{\frac{2}{3}}r_S^{\frac{1}{3}}$. Redshift diverges at ${r_E}
=r_S$ due to this fact that there is an event horizon at this radii.
There is also another singularity at $(\bar r - \bar t)_E=0$ in
accordance with the coordinate singularity of metric at the origin
($r=0$).

By defining the conformal time $\bar t$ as
\begin{eqnarray}\label{newtime}
\bar t=\int_{0}^{t}\frac{dt'}{a(t')}.
\end{eqnarray}
one can rewrite Eq.~(\ref{FRW}) as
\begin{eqnarray}\label{cFRW}
ds^2=a(\bar t)^2(-d\bar t^2+[\frac{dr^2}{1-kr^2}+r^2d\Omega^2]),
\end{eqnarray}
where $a(\bar t)=\frac{A^2}{2}\bar t$ and $a(\bar
t)=\frac{A^3}{9}\bar t^2$ for $\omega=\frac{1}{3}$ and the matter
domination era ($\omega=0$) respectively. We have also $a(\bar
t)=\frac{A}{1-AH\bar t}$ for the mysterious dark energy era
($\omega=-1$). Indeed, the speed of light, as the traversed distance
in the unit conformal time, is one in this new coordinates.
Therefore, the universe expansion eras can also be modeled by using
conformal scale factor $a(\bar t)$ instead of $a(t)$.

Considering the conformal scale factor $a(\bar t)$, we introduce metric
\begin{equation}\label{conformal Lemaitre}
dS_{CL}^2 = a{(\bar t)^2}\left( {d{{\bar t}^2} -
\frac{{{r_S}}}{r}d{{\bar r}^2}-{r^2}d{\Omega ^2}} \right),
\end{equation}
which is the conformal transformation of metric (\ref{Lemaitre
metric}) and therefore, its corresponding causal structure is the
same as that of (\ref{Lemaitre metric}) \cite{wald} which can be
found in \cite{Shank}. This metric suffers generally from two
singularities at $a(\bar t)=0$ (big bang) and $r=0$ (origin). Simple
calculation for the redshift leads to
\begin{equation}\label{Conformal Lemaitre Redshift}
1 + z = \frac{{{\lambda _O}}}{{{\lambda _E}}} = \frac{{a({{\bar
t}_O})}}{{a({{\bar t}_E})}} \times \sqrt {\frac{{1 - {{\left(
{\frac{{2{r_S}}}{{3{{(\bar r - \bar t)}_O}}}} \right)}^{2/3}}}}{{1
- {{\left( {\frac{{2{r_S}}}{{3{{(\bar r - \bar t)}_E}}}}
\right)}^{2/3}}}}}.
\end{equation}
Redshift diverges at $\frac{3}{2}(\bar r - \bar t)_E=r_S$ and
$\frac{3}{2}(\bar r - \bar t)_E=0$ the same as the Lemaitre
metric. In addition, the FRW result is covered in the $r_S=0$
limit as a desired result.
\subsection*{Curvature scalars and energy conditions}
Let us calculate the surface area of a hypersurface with constant
radii $r$ at constant time $\bar t$
\begin{eqnarray}\label{area}
A=\int a(\bar t)^2 r^2 d\Omega=4\pi a(\bar t)^2 r^2,
\end{eqnarray}
where $r$ was defined in Eq. (\ref{r}). Therefore, we have $A=0$ for
$r=0$ and $A\neq0$ for $r=r_S$ signalling that there is a curvature
singularity at $r=0$. For the normal to the hypersurface,
$\phi=(\frac{3}{2}(\bar r - \bar t))^{\frac{2}{3}}r_S^{\frac{1}{3}}
-C=0$, we have
\begin{eqnarray}\label{normal}
n_{\alpha}=(-1,1,0,0)(\bar r - \bar t)^{\frac{-1}{3}}r_S^{\frac{1}{3}}.
\end{eqnarray}
leading to
\begin{eqnarray}\label{normal1}
n_{\alpha}n^{\alpha}=(\bar r - \bar t)^{\frac{-2}{3}}r_S^{\frac{2}{3}}
a(\bar t)^{-2}(1-\frac{r}{r_S}).
\end{eqnarray}
Therefore, $r=r_S$ is a null hypersurface which is the same as that
of~(\ref{Lemaitre metric}). In order to clarify the nature of the
redshift singularities, we use the curvature scalars. Calculations
lead to
\begin{equation}\label{Ricci Scalar}
R = \frac{{6\left( {\dot a - \ddot a(\bar r - \bar t)}
\right)}}{{{a^3}(\bar r - \bar t)}},
\end{equation}
and
\begin{equation}\label{Ricci square}
\begin{array}{l}
{R^{\mu \nu }}{R_{\mu \nu }} = \frac{{12{{\ddot a}^2}}}{{{a^6}}} -
\frac{{\dot a\ddot a\left( {12\dot a(\bar r - \bar t)+ 16a} \right)}}{{{a^7}(\bar r - \bar t)}}\\
\,\,\,\,\,\,\,\,\,\,\,\,\,\,\,\,\,\,\,\,\,\,\, + \frac{{{{\dot
a}^2}\left( {12{{\dot a}^2}{{(\bar r - \bar t)}^2} - 4a\dot a(\bar
r - \bar t) + 12{a^2}} \right)}}{{{a^8}{{(\bar r - \bar t)}^2}}},
\end{array}
\end{equation}
for the Ricci scalar and square. Furthermore, for the Rieman and
Weyl squares we get
\begin{equation}\label{Kretchmann Scalar}
\begin{array}{l}
K = \frac{{24{{\dot a}^4}}}{{{a^8}}} + \frac{{12{{\dot a}^2} - 8\dot a\ddot a(\bar r - \bar t) + 12{{\ddot a}^2}{{(\bar r - \bar t)}^2}}}{{{a^6}{{(\bar r - \bar t)}^2}}}\\
\,\,\,\,\,\,\,\,\,\,\, - \frac{{8{{\dot a}^3} + 24{{\dot a}^2}\ddot a(\bar r - \bar t)}}{{{a^7}(\bar r - \bar t)}} + \frac{{64}}{{27}}\frac{1}{{{a^4}{{(\bar r - \bar t)}^4}}}\,\\
\,\,\,\,\,\,\,\,\,\,
\end{array}
\end{equation}
and
\begin{equation}\label{Weyl invariant}
W = \frac{{64}}{{27{a^4}{{(\bar r - \bar t)}^4}}}.
\end{equation}
It can be seen that, unlike $r=0$, none of them diverge at $r=r_S$.
Therefore, $r=0$ is a naked singularity while $r=r_S$ indicates that
we face with an event horizon at this radii. Here, we can conclude
that $r=r_S$ is the co-moving radii of the event horizon while its
physical radii ($\xi$) is $\xi=a(\bar t)r_S$ in accordance with the
FRW background. Therefore this event horizon covers the origin
singularity ($r=0$). Finally, we should note that the FRW results
are accessible in the appropriate limit ($(\bar r - \bar t)\gg1$).
Bearing the Einstein field equation in mind, the energy-momentum
tensor supporting the conformal Lemaitre metric~(\ref{conformal
Lemaitre}) is as follows,
\begin{equation}\label{energy-momentum tensor}
{T^\mu }_\nu  = \left( {\begin{array}{*{20}{c}}
\rho &0&0&0\\
0&{{p_{11}}}&0&0\\
0&0&{{p_{22}}}&0\\
0&0&0&{{p_{33}}}
\end{array}} \right)
\end{equation}
where $\rho$ is the energy density and  $p_{11}$ and $p_{22}(=p_{33})$ are the pressure components.
The energy density and pressure components are obtained as follows
\begin{equation}\label{Energy density}
\rho=\frac{{3{{\dot a}^2}(\bar r - \bar t) - 2a\dot
a}}{{{a^4}(\bar r - \bar t)}},
\end{equation}
\begin{equation}\label{Pressure T11}
{p_{11}} = \frac{{{{\dot a}^2}(\bar r - \bar t) + \frac{8}{3}a\dot a -
2a\ddot a(\bar r - \bar t)}}{{{a^4}(\bar r - \bar t)}},
\end{equation}
\begin{equation}\label{Pressure T22}
{p_{22}} ={p_{33}}= \frac{{{{\dot a}^2}(\bar r - \bar t) +
\frac{2}{3}a\dot a - 2a\ddot a(\bar r - \bar t)}}{{{a^4}(\bar r - \bar
t)}}.
\end{equation}
It is be mentioned that, in the $r\gg1$ limit ($(\bar r - \bar
t)\gg1$), these relations converge to those of FRW as a desirable
expectation. The weak (WEC) and strong (SEC) energy conditions state
that
\begin{equation}\label{WEC}
\begin{array}{l}
\rho  \ge 0\\
\rho  + {p_{ii}} \ge 0,
\end{array}
\end{equation}
and
\begin{equation}\label{SEC}
\begin{array}{l}
\rho  + {p_{ii}} \ge 0\\
\rho  + \sum\limits_{i = 1}^3 {{p_{ii}}}  \ge 0,
\end{array}
\end{equation}
respectively. Combining Eqs.(\ref{r}) and (\ref{Energy density}) and
using WEC, we see that the $\rho\geq0$ is valid when radii meets the
$r\geq(\frac{r_S}{h^2})^{\frac{1}{3}}\equiv r_{\rho}$ condition. It
should be noted that $h\equiv\frac{\dot{a}(\bar t)}{a(\bar t)}$ and
($\dot{}$) denotes derivative with respect $\bar t$. In order to
satisfy $\rho+{p_{22}}\geq0$, by combining Eqs.~(\ref{r})
and~(\ref{Pressure T22}), we get $r\geq(\frac{h^2r_S}
{(\dot{h}-h^2)^2})^{\frac{1}{3}}\equiv r_r$. Since $a(\bar t)\sim
\bar t$ and $a(\bar t)\sim \bar t^2$ for the radiation (RDA) and
matter (MDA) dominated eras respectively, $h>0$, $\dot{h}<0$ and
thus $(h^2-\dot{h})>h^2$ yielding $r_{\rho}>r_r$. Therefore, when
$r$ meets the $r\geq r_\rho$ condition, the $\rho\geq0$ and
$\rho+{p_{22}}\geq0$ conditions are satisfied, in these regimes,
simultaneously. The condition $\rho+{p_{11}}\geq0$ leads to
$\frac{2r^{\frac{3}{2}}}{r_S^{\frac{1}{2}}}(h^2-\dot{h})+h\geq0$
which is valid in the RDA and MDA regimes ($\dot{h}<0$). Moreover,
the $\rho+\sum\limits_{i = 1}^3 {{p_{ii}}}\geq0$ condition leads to
$\frac{r^{\frac{3}{2}}}{r_S^{\frac{1}{2}}}(-\dot{h})+\frac{h}{3}\geq0$
which is again valid in the RDA and MDA regimes. Therefore,
independent of $\bar t$, the energy conditions are satisfied when
$r$ meets the $r\geq r_\rho$ condition in the the RDA and MDA eras.
Finally, we should note that since in the FRW limit
$r_S\rightarrow0$, $r_\rho\rightarrow0$ and therefore, the energy
conditions will be fully satisfied.

Using the conformal scale factor for the mysterious dark energy era
introduced in previous section, since $H$ is constant in the
standard cosmology including a fluid with non-negative density and
$\omega=-1$ for describing dark energy called cosmological constant
\cite{mukh,wein}, we have $\dot{h}=h^2>0$. Therefore, $\rho\geq0$
yields $r\geq r_\rho$.In this era, $\rho+{p_{22}}\geq0$ leads to
$-2h>0$ indicating that this condition is not valid in this regime.
In addition, $\rho+{p_{11}}\geq0$ leads to $h\geq0$ which is valid
independent of $r$ and $\bar t$. If $r$ meet the
$r\leq(\frac{r_S}{4h^2})^\frac{1}{3}=\frac{r_\rho}{4^\frac{1}{3}}$
condition, then the $\rho+\sum\limits_{i = 1}^3 {{p_{ii}}}\geq0$
condition will be also valid. Therefore, for $r\geq r_\rho$, only
the $\rho\geq0$ and $\rho+{p_{11}}\geq0$ conditions are satisfied.
Here, we must note that since the values of $\rho$ and $p_{ii}$
converge to those of FRW in the $(\bar r - \bar t)\gg1$ limit, the
energy conditions are asymptotically converged to their values in
the FRW metric where $\rho>0$, $\rho+{p_{ii}}=0$ and
$\rho+\sum\limits_{i = 1}^3 {{p_{ii}}}\leq0$ are valid for a
de-Sitter universe ($\omega=-1$) \cite{wein,mukh}. The same result
is valid in the $r_S\rightarrow0$ limit.
\section{Conformally outward Lemaitre Spacetime}
In the previous section, we studied the mathematical and physical
features of introduced metric which is conformal to the Lemaitre
metric for a system in which the particles freely falling into a
gravitational field due to a point-like mass $M$. Here, we extend
our analysis to the metric which is conformal to the Lemaitre metric
describing a system where particles trajectories move outward from
the singularity at $r=0$. To do this, we should choose minus sign
in~(\ref{Lemaitre transformations}), and we will get
metric~(\ref{Lemaitre metric}). But, here, we have
\begin{equation}\label{r1}
r = {\left( {\frac{3}{2}(\bar r + \bar t)}
\right)^{\frac{2}{3}}}r_S^{\frac{1}{3}}.
\end{equation}
Using conformal transformation and after some algebra, we get:
\begin{equation}\label{Conformal Lemaitre Redshift1}
1 + z = \frac{{{\lambda _O}}}{{{\lambda _E}}} = \frac{{a({{\bar
t}_O})}}{{a({{\bar t}_E})}} \times \sqrt {\frac{{1 - {{\left(
{\frac{{2{r_S}}}{{3{{(\bar r + \bar t)}_O}}}} \right)}^{2/3}}}}{{1
- {{\left( {\frac{{2{r_S}}}{{3{{(\bar r + \bar t)}_E}}}}
\right)}^{2/3}}}}}.
\end{equation}
Again, we see that redshift diverges at $(\bar r + \bar t)=0$ and
$(\bar r + \bar t)=r_S$ while the surface area at this radiuses are
the same as those of obtained in the previous section. In addition,
for a hypersurface with $r=C$ we find
\begin{eqnarray}\label{normal2}
n_{\alpha}n^{\alpha}=(\bar r + \bar t)^{\frac{-2}{3}}r_S^{\frac{2}{3}}
a(\bar t)^{-2}(1-\frac{r}{r_S}),
\end{eqnarray}
indicating that we have a null hypersurface at $r=r_S$ the same as
the previous case. We have also
\begin{equation}\label{Ricci Scalar1}
R = -\frac{{6\left( {\dot a + \ddot a(\bar r + \bar t)}
\right)}}{{{a^3}(\bar r + \bar t)}},
\end{equation}
and
\begin{equation}\label{Ricci square1}
\begin{array}{l}
{R^{\mu \nu }}{R_{\mu \nu }} = \frac{{12{{\ddot a}^2}}}{{{a^6}}} +
\frac{{\dot a\ddot a\left(16a-{12\dot a(\bar r + \bar t)} \right)}}{{{a^7}(\bar r + \bar t)}}\\
\,\,\,\,\,\,\,\,\,\,\,\,\,\,\,\,\,\,\,\,\,\,\, + \frac{{{{\dot
a}^2}\left( {12{{\dot a}^2}{{(\bar r + \bar t)}^2} + 4a\dot a(\bar
r + \bar t) + 12{a^2}} \right)}}{{{a^8}{{(\bar r + \bar t)}^2}}},
\end{array}
\end{equation}
for the Ricci scalar and square. Moreover,
\begin{equation}\label{Kretchmann Scalar1}
\begin{array}{l}
K = \frac{{24{{\dot a}^4}}}{{{a^8}}} + \frac{{12{{\dot a}^2} + 8\dot a\ddot a(\bar r + \bar t) + 12{{\ddot a}^2}{{(\bar r + \bar t)}^2}}}{{{a^6}{{(\bar r + \bar t)}^2}}}\\
\,\,\,\,\,\,\,\,\,\,\, +\frac{{8{{\dot a}^3} - 24{{\dot a}^2}\ddot a(\bar r + \bar t)}}{{{a^7}(\bar r + \bar t)}} + \frac{{64}}{{27}}\frac{1}{{{a^4}{{(\bar r + \bar t)}^4}}}\,\\
\,\,\,\,\,\,\,\,\,\,
\end{array}
\end{equation}
and
\begin{equation}\label{Weyl invariant1}
W = \frac{{64}}{{27{a^4}{{(\bar r + \bar t)}^4}}}.
\end{equation}
are the Rieman and Weyl squares respectively. As again, none of them diverge at
$r=r_S$. In addition, we note that since our spacetime is conformal to the Lemaitre,
its causal structure is the same that of Lemaitre. Therefore, exactly the same as the
previous case, $r=0$ and $r=r_S$ point to a naked singularity and the co-moving radii
of an event horizon respectively, while for the physical radii we have $\xi=a(\bar t)r$.
This spacetime implies
\begin{equation}\label{Energy density1}
\rho=\frac{{3{{\dot a}^2}(\bar r + \bar t) + 2a\dot
a}}{{{a^4}(\bar r + \bar t)}},
\end{equation}
\begin{equation}\label{Pressure T111}
{p_{11}} = \frac{{{{\dot a}^2}(\bar r + \bar t) + \frac{8}{3}a\dot a -
2a\ddot a(\bar r + \bar t)}}{{{a^4}(\bar r + \bar t)}},
\end{equation}
\begin{equation}\label{Pressure T221}
{p_{22}} ={p_{33}}= \frac{{{{\dot a}^2}(\bar r + \bar t) +
\frac{2}{3}a\dot a - 2a\ddot a(\bar r + \bar t)}}{{{a^4}(\bar r + \bar
t)}}.
\end{equation}
It can be found that in the $(\bar r + \bar t)\gg1$ limit, the
results of the FRW metric are obtainable. In addition, $\rho\geq0$
is valid everywhere and it is independent of the conformal scale
factor ($a(\bar t)$). $\rho+{p_{11}}\geq0$ leads to $3(\bar r + \bar
t)[h^2-\dot{h}]+7h\geq0$ which, independent of the expansion regime,
it is valid . For $\rho+{p_{22}}\geq0$, we get $3(\bar r + \bar
t)[h^2-\dot{h}]+4h\geq0$ indicating that this condition is again
valid in the RDA and MDA regimes as well as the dark energy era.
Finally, we can show that the $\rho+{p_{11}}+2{p_{22}}\geq0$
condition leads to $3(\bar r + \bar t)[-\dot{h}]+h\geq0$. The latter
is valid in the RDA as well as MDA regimes, since $\dot{h}<0$. In
order to satisfy $\rho+\sum\limits_{i = 1}^3 {{p_{ii}}}\geq0$ in the
dark energy era, $r$ should meet the
$r\leq(\frac{9r_S}{4h^2})^{\frac{1}{3}}$ condition. In summary, the
validity of the energy conditions in the RDA as well as MDA regimes
is independent of $r_S$, while it depends on the mass enclosed by
the event horizon located at $r_S$ in the dark energy era. Loosely
speaking, the energy conditions are not valid in the dark energy era
for $r>(\frac{9r_S}{4h^2})^{\frac{1}{3}}$.
\section{Conclusions}
Since the Lemaitre transformations of the Schwarzschild metric
provides the suitable metrics for studying the freely moving
particles in a gravitational field due to the massive motionless
object ($M$), we used this metric to encounter the effects of the
universe expansion on the horizon. Indeed, the singularity of the
Schwarzschild metric at the horizon radii is eliminated by Lemaitre
transformations leading to this fact that if one embedded this
metric into the FRW background, using the conformal transformation,
then the resultant metric and its physical and mathematical
properties will be well-defined at this radii. We should note again
that since we have conformal transformed the Lemaitre metrics, the
causal structure of the resultant metrics are the same as the
primary metrics. In addition, the physical radii of the horizon is
increased as a function of the universe expansion. Energy conditions
are investigated for the conformal incoming and oncoming Lemaitre
metrics. Therefore, unlike other works mentioned in the
introduction, the conformal transformation of the Lemaitre metrics
can provide suitable spacetimes for investigating the effects of the
universe expansion on the local physics such as the BHs and thus the
dynamic Black Holes.
\acknowledgments{This work has been supported financially by
Research Institute for Astronomy and Astrophysics of Maragha, Iran.}

\end{document}